\documentclass{phb-proc4-auth}
\usepackage{graphicx}
\usepackage{amssymb}
\def \virg{,}
\def \point{.}
\def \e{{\rm e}}
\def \h{{\rm h}}
\def \TC{T_{\rm C}}
\def \EB{E_{\rm B}}
\renewcommand{\Im}{\mathop{\rm Im}\nolimits}
\def\jo #1#2#3#4{#1 #2 (#3) #4}   %For Elsevier Science
\def\PRB{Phys.\ Rev.\ B}
\def\PRL{Phys.\ Rev.\ Lett.}

\def\RMP{Rev.\ Mod.\ Phys.}

\def\JLTP{J.\ Low\ Temp.\ Phys.}
\def\RMP{Rev.\ Mod.\ Phys.}
\begin{document}
\begin{frontmatter}

\journal{EXCON '04 Proceedings}
\date{30 June 2004}

\title{\bf Phase diagram for the exciton Mott transition 
in infinite-dimensional electron-hole systems}

\author{Yuh Tomio\corauthref{cor1}}, 
\ead{tomio@acty.phys.sci.osaka-u.ac.jp} 
\author{Tetsuo Ogawa}

\corauth[cor1]{Corresponding author. 
Tel.: +81-6-6850-5735; fax: +81-6-6850-5351}

\address{CREST, JST and Department of Physics, Osaka University,  
 Toyonaka, Osaka 560-0043, Japan}

\begin{abstract}
To understand the essence of the exciton Mott transition 
in three-dimensional electron-hole systems, 
the metal-insulator transition is studied 
for a two-band Hubbard model in infinite dimensions 
with interactions of 
electron-electron (hole-hole) repulsion $U$ 
and electron-hole attraction $-U'$. 
By using the dynamical mean-field theory, 
the phase diagram in the $U$-$U'$ plane is obtained 
 (which is exact in infinite dimensions) 
 assuming that electron-hole pairs do not condense. 
When both electron and hole bands are half-filled, 
two types of insulating states appear: 
the Mott-Hubbard insulator for $U > U'$ and 
the biexciton-like insulator for $U < U'$.   
Even when away from half-filling, we find 
the phase transition between the exciton- or biexciton-like 
insulator and a metallic state. 
This transition can be assigned to the exciton Mott transition,   
whereas the Mott-Hubbard transition is absent. 
\end{abstract}

\end{frontmatter}

% main text
%%%%%%%%%%%%%%%%%%%%%%%%%%%%%%%%%%%%%%%%%%%%%%%%%%%%%
\section{Introduction}

Electron-hole (e-h) systems in photoexcited semiconductors 
exhibit various remarkable properties depending on carrier density, 
temperature, etc., and have been investigated extensively both 
experimentally and theoretically~\cite{Moskalenko00}. 
In particular, the metal-insulator transitions 
have attracted interest for many years: 
the exciton Mott transition at high temperatures between 
an exciton/biexciton gas phase and an e-h plasma phase, and 
crossover at low temperatures between the Bose-Einstein condensation
(BEC) of excitons at low density and the BCS-like condensation of 
e-h pairs at high density. 
However, 
the complicated tangle of the elements, two types of fermions,  
 Coulomb interactions of both repulsion and attraction, 
  screening effects, e-h densities, temperatures, etc., 
 makes the physics of this system hard to be understood.   
Therefore,  theoretical understanding of especially 
the exciton Mott transition and the BEC-BCS crossover 
is still not sufficient.

We expect that a study from a standpoint of the strong-correlation
physics provides new interpretation about the above problems.  
As the  first step of our work, we examine 
the exciton Mott transition in consideration of the minimum elements, 
i.e., a two-band Hubbard model,  
by using the dynamical mean-field theory (DMFT)~\cite{Georges96} 
recently developed through the study of 
strongly correlated electron systems. 
The DMFT requires only the locality of the self-energy,  
and can take full account of local correlations. 
This locality and the resulting DMFT 
become exact in the limit of infinite spatial dimensions 
and  good approximation of the three-dimensional systems.  

In the present paper, we focus on the normal phase 
where the condensation of e-h pairs (i.e., exciton BEC and 
e-h BCS state) is not allowed.   
The calculation is performed at absolute zero temperature. 

%%%%%%%%%%%%%%%%%%%%%%%%%%%%%%%%%%%%%%%%%%%%%%%%%%%%%
\section{Two-site dynamical mean-field theory}

 We consider a electron-hole system  
described by the two-band Hubbard model given by  
%============ eq.(1) ====================
\begin{eqnarray}
 \label{H}
 H &=& -\sum_{<ij>\sigma}\sum_{\alpha=\e,\h}
     t_\alpha d_{i\sigma}^{\alpha\dagger} d_{j\sigma}^\alpha
    -  \sum_{i\sigma,\alpha} \mu_\alpha
     d_{i\sigma}^{\alpha\dagger} d_{i\sigma}^\alpha  
\nonumber \\
&& {}
 + U \sum_{i,\alpha} 
d_{i\uparrow}^{\alpha\dagger} d_{i\uparrow}^\alpha
d_{i\downarrow}^{\alpha\dagger} d_{i\downarrow}^\alpha
- U' \sum_{i\sigma\sigma'} 
d_{i\sigma}^{\e\dagger} d_{i\sigma}^\e
d_{i\sigma'}^{\h\dagger} d_{i\sigma'}^\h
\virg 
\end{eqnarray}
%========================================
where $d^{\e\dagger}_{i\sigma}$ ($d^{\h\dagger}_{i\sigma}$) 
 denotes a creation operator of 
a conduction band electron (a valence band hole)  
with spin $\sigma=\{\uparrow,\downarrow\}$ at the $i$-th site. 
The quantities $t_\e$ ($t_\h$) and $\mu_\e$ ($\mu_\h$) are  
the transfer integral of the electrons (holes) 
between the neighboring sites and 
the chemical potential measured from the center of 
the electron (hole) band, respectively. 
The on-site Coulomb interaction 
of the e-e (h-h) repulsion and 
that of the e-h attraction 
are expressed by $U$ and $-U'$, respectively. 
The local Green function for electrons or holes
of the model~(\ref{H}) is defined by
%============ eq.(2) ====================
\begin{equation}
 \label{G}
G^\alpha(\omega)=
 \langle\!\langle 
d_{i\sigma}^\alpha;d_{i\sigma}^{\alpha\dagger}
 \rangle\!\rangle_\omega
=\int^\infty_{-\infty} d\varepsilon 
\frac{\rho_0^\alpha(\varepsilon)}
{\omega+\mu_\alpha-\varepsilon-\Sigma^\alpha(\omega)}
\virg
\end{equation}
%========================================
where $\Sigma^\alpha(\omega)$ is the self-energy of 
electrons ($\alpha=\e$) or holes ($\alpha=\h$), which is local, 
i.e., does not depend on the wave number, 
in the limit of infinite-dimensions. 
We use the semicircular density of states (DOS),  
$\rho_0^\alpha(\varepsilon)=
\sqrt{4t_\alpha^2-\varepsilon^2}/(2\pi t_\alpha^2)
$.

Within the DMFT~\cite{Georges96}, the many-body problem of 
the lattice model~(\ref{H}) is mapped onto the problem of 
a single-site impurity embedded in an effective medium.  
The effective medium, which is dynamical and 
is represented by the noninteracting impurity Green function 
$\mathcal{G}_0^\alpha(\omega)$ of an effective 
single-impurity Anderson model (SIAM), 
is determined from the self-consistency condition 
$\mathcal{G}_0^\alpha(\omega)^{-1}=\omega+\mu_\alpha
-t_\alpha^2 G^\alpha(\omega)$.  
The condition is also read as 
$G^\alpha_{\rm imp}(\omega)=G^\alpha(\omega)$. 
The interacting impurity Green function 
of the effective SIAM, $G^\alpha_{\rm imp}(\omega)$, 
should be calculated exactly such that effects of 
the interactions on the impurity site are fully included.  
Contrary to the ordinary mean-field approaches, 
thus, in the DMFT scheme the local correlations and 
dynamical quantum fluctuations are taken into full account. 

In order to extract a sketch of the phase diagram of 
the model~(\ref{H}),  
here we apply the two-site DMFT~\cite{Potthoff01} 
which is simplified version of the DMFT. 
In the two-site DMFT, 
the effective medium $\mathcal{G}_0^\alpha(\omega)$ is 
represented approximately by only the fewest parameters, 
i.e., the effective SIAM  consists of 
a single impurity and only a single bath sites. 
Since the essence of the DMFT concerning the local correlations 
still remains despite the bold approximation and simplification,  
it can be successful to provide 
the most correct critical point of the Mott-Hubbard transition 
and to describe the qualitative 
electronic properties~\cite{Potthoff01}.

For the model~(\ref{H}), 
the corresponding effective two-site SIAM is written as 
%============ eq.(3) ====================
\begin{eqnarray}
 \label{Himp}
 H_{\rm imp} &=& 
 \sum_{\sigma,\alpha} \left[ 
\varepsilon_c^{\alpha} c_{\sigma}^{\alpha\dagger} c_{\sigma}^{\alpha}
+  V_\alpha \left(  
d_{\sigma}^{\alpha\dagger} c_{\sigma}^{\alpha}
 + {\rm h.c.} \right) 
 -  \mu_{\alpha} 
d_{\sigma}^{\alpha\dagger} d_{\sigma}^{\alpha} \right]
\nonumber \\
&& {}
 + U \sum_{\alpha} 
d_{\uparrow}^{\alpha\dagger} d_{\uparrow}^\alpha
d_{\downarrow}^{\alpha\dagger} d_{\downarrow}^\alpha
- U' \sum_{\sigma\sigma'} 
d_{\sigma}^{\e\dagger} d_{\sigma}^\e
d_{\sigma'}^{\h\dagger} d_{\sigma'}^\h
\virg
\end{eqnarray}
%========================================
where the bath parameters $V_\alpha$ and $\varepsilon_c^\alpha$ 
denote the hybridization between the impurity ($d$) and 
bath ($c$) sites, and the energy level of the bath site, respectively.  
 The Green function of the effective medium 
(i.e., noninteracting impurity Green function) becomes 
$\mathcal{G}_0^\alpha(\omega)^{-1}=
\omega+\mu_\alpha-V_\alpha^2/(\omega-\varepsilon_c^\alpha)$. 
In the two-site DMFT, the self-consistency condition is reduced to 
simpler equation by the following procedure~\cite{Potthoff01}:  
the self-energy is expanded in the low-energy region, 
$\Sigma^\alpha(\omega) \sim a_\alpha+b_\alpha\omega$, and then 
 the resulting local Green function~(\ref{G}) and 
impurity Green function $G^\alpha_{\rm imp}(\omega)^{-1}
=\mathcal{G}_0^\alpha(\omega)^{-1}-\Sigma^\alpha(\omega)$
are compared so as to coincide at high-energy region.  
Thereby, the self-consistency equation for $V_\alpha$ 
is obtained as  
%============ eq.(4) ====================
\begin{equation}
\label{SCEQ-V}
 V_\alpha^2 = t_\alpha^2 Z_\alpha
\virg
\end{equation}
%========================================
where 
%============ eq.(5) ====================
\begin{equation}
\label{Z}
Z_\alpha=\left(1-b_\alpha\right)^{-1}=
\left[1-\frac{d\Sigma^\alpha(\omega)}{d\omega}\big|_{\omega=0}\right]^{-1}
\virg
\end{equation}
%========================================
 is the quasiparticle weight which generally characterizes 
the Fermi liquid (metallic) states. 
On the other hand, the requirement that the particle densities 
of the original and impurity models must be equal, i.e., 
$n^\alpha=n^\alpha_{\rm imp}$, 
leads to the self-consistency condition for $\varepsilon_c^\alpha$, 
%============ eq.(6) ====================
\begin{equation}
\label{SCEQ-E}
\int^0_{-\infty} d\omega \Im{G^\alpha(\omega+i0^+)} =
\int^0_{-\infty} d\omega \Im{G^\alpha_{\rm imp}(\omega+i0^+)} 
\point
\end{equation}
%========================================

Consequently, the model~(\ref{H}) can be solved 
within the two-site DMFT by the following self-consistency cycle: 
(i)~$G^\alpha_{\rm imp}(\omega)$ is directly calculated by 
the exact diagonalization of the two-site SIAM~(\ref{Himp}) 
with $\varepsilon_c^\alpha$ and $V_\alpha$. 
(ii)~By using $\Sigma^\alpha(\omega)=
\mathcal{G}_0^\alpha(\omega)^{-1}-G^\alpha_{\rm imp}(\omega)^{-1}$,   
a new value of $V_\alpha$ is determined from 
the condition~(\ref{SCEQ-V}) and Eq.~(\ref{Z}). 
(iii)~By substituting  $\Sigma^\alpha(\omega)$ for Eq.~(\ref{G}), 
a new value of $\varepsilon_c^\alpha$ is chosen so as to satisfy 
the condition~(\ref{SCEQ-E}). 
This process (i)-(iii) is iterated until 
$\varepsilon_c^\alpha$ and $V_\alpha$ converge.

The metal-insulator transition for the normal phase 
of the model~(\ref{H}) is discussed from  behaviors of 
both the quasiparticle weight $Z_\alpha$ and 
the interacting DOS 
$\rho^\alpha(\omega)=-\Im G^\alpha(\omega+i0^+)/\pi$, 
with varying $U$, $U'$, $t_\h/t_\e$ and also 
the particle density $n$ $(\equiv n^\e=n^h)$.

%%%%%%%%%%%%%%%%%%%%%%%%%%%%%%%%%%%%%%%%%%%%%%%%%%%%%
\section{Phase diagram at half filling}

First, we concentrate on the special case where 
the both electron and hole bands are half-filled, i.e., $n=1$. 
In this symmetric case, we can set $\mu_\alpha=U/2-U'$ 
and $\varepsilon_c^\alpha=0$.

\begin{figure}[b]
 \includegraphics[height=4.9cm,clip,keepaspectratio]{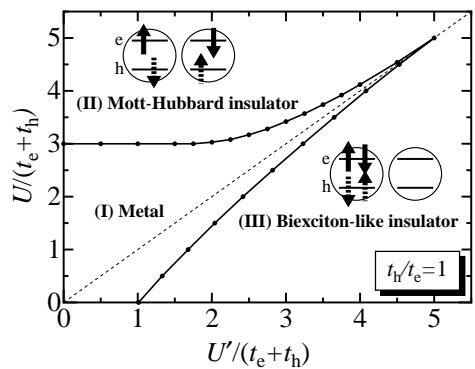}
\caption{%
Phase diagram in the $U'$-$U$ plane 
at half-filling ($n=1$) for $t_\h/t_\e=1$. 
}%
\label{Fig1}
\end{figure}

%------------- figure 1 ----------------------
For $t_\h/t_\e=1$ 
(the mass of the hole is the same as that of the electron), 
the phase diagram on the plane of $U'$ and $U$ is 
 shown in  Fig.~\ref{Fig1}. 
There are three kinds of states: (I) metallic state, 
(II) Mott-Hubbard insulating state, 
and (III) biexciton-like insulating state. 
The second-order transitions between these states 
occur on the solid curves. 
In the metallic state (I), $Z_\alpha$ has a finite value and 
there is finite DOS at the Fermi level
(the quasiparticle coherent peak), i.e., $\rho^\alpha(0) \neq 0$.   
On the other hand, in the both insulating states (II) and (III),  
$Z_\alpha=0$  and the coherent peak of the DOS disappears. 
However, the physical pictures of the insulating 
states (II) and (III)
are quite different, as drawn schematically in Fig.~\ref{Fig1}:  
the state (II) is induced by the e-e (h-h) repulsion $U$ on each 
electron and hole band,  
while the state (III) is realized by the e-h attraction $U'$ 
on each site. 
The competition of these two states stabilizes
the metallic state for $U \simeq U'$.

We point out that 
the above results are equivalent to those obtained for 
the two-orbital repulsive Hubbard model~\cite{Koga02}   
because this model and our model~(\ref{H}) 
only at half-filling can be mapped onto each other 
by the attraction-repulsion transformation. 
Actually, the phase diagram of Fig.~\ref{Fig1} 
is in good agreement with that of Ref.~\cite{Koga02}.

%************** fig.(1) *******************
%\begin{center}
%\end{center}
%******************************************

\begin{figure}[tb]
 \includegraphics[height=4.9cm,clip,keepaspectratio]{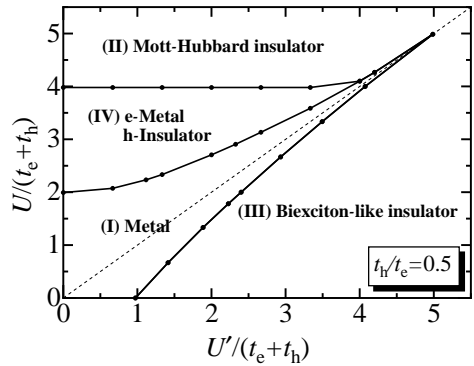}
\caption{%
Phase diagram in the $U'$-$U$ plane 
at half-filling ($n=1$) for $t_\h/t_\e=0.5$.
}%
\label{Fig2}
\end{figure}

%------------- figure 2 ----------------------
We also examine effects of the difference between  
 the electron and hole masses. 
In  Fig.~\ref{Fig2}, 
the phase diagram on the plane of $U'$ and $U$ is shown  
for $t_\h/t_\e=0.5$, 
 where the hole is twice as heavy as the electron. 
A new state (IV) appears between states (I) and (II),   
in which $Z_\e \neq 0$ but $Z_\h=0$, i.e.,  
the electron (hole) band is metallic (insulating). 
In other words, 
the Mott-Hubbard transition of holes
  does not coincide with that of electrons when $t_\e \neq t_\h$. 
 This ``band-selective'' Mott-Hubbard transition  
 corresponds to the ``orbital-selective'' Mott transition 
in the two-orbital repulsive model~\cite{Koga04}. 
But more than the above, what we should emphasize here 
from common features of Figs.~\ref{Fig1} and \ref{Fig2}  
is as follows. 
(i)~The metal-insulator transition between states (I) and (III) 
 is by no means ``band-selective'' for any  ratio  $t_\h/t_\e$. 
(ii)~The position of that phase boundary  
on the plane of interactions scaled by $t_\e+t_\h$  
is universal  with regard to the ratio $t_\h/t_\e$.  
These facts  
indicate that the transition between the metallic state (I) and 
the biexciton-like insulator (III) occurs as a result 
of the competition between the interactions 
and the relative motion of electron and hole.   
Note that the quantity $t_\e+t_\h$ is proportional to the 
energy of the relative motion.  

%************** fig.(2) *******************
%\begin{center}
%\end{center}
%******************************************

%%%%%%%%%%%%%%%%%%%%%%%%%%%%%%%%%%%%%%%%%%%%%%%%%%%%%
\section{Phase diagram at arbitrary filling}

In this section, we discuss the case of arbitrary filling.
For $n \neq 1$, the process for determining of 
the chemical potential $\mu_\alpha$ is added to 
the self-consistency cycle for $\varepsilon_c^\alpha$ and $V_\alpha$.
Hereafter $t_\h/t_\e=1$ is fixed.

\begin{figure}[tb]
 \includegraphics[height=4.9cm,clip,keepaspectratio]{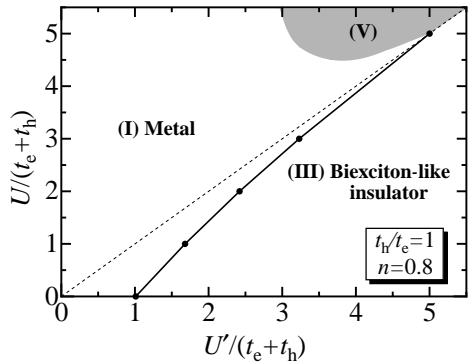}
\caption{%
Phase diagram in the $U'$-$U$ plane for $t_\h/t_\e=1$ and $n=0.8$.
 The shaded area indicates the region speculated that  
the exciton-like insulator  
(local e-h pairing state) appears. 
}%
\label{Fig3}
\end{figure}

%------------- figure 3 ----------------------
Fig.~\ref{Fig3} shows the phase diagram on the plane of  
$U'$ and $U$ for $n=0.8$. 
The Mott-Hubbard insulator (II) disappears immediately 
away from half-filling, 
as known in the single-band Hubbard model~\cite{Georges96,Potthoff01}, 
while the metallic state (I) and 
the biexciton-like insulator (III) remain.   
In the present two-site DMFT calculation,  
only the metallic state in which  
both $Z_\alpha$ and $\rho^\alpha(0)$ are nonzeroes, 
is always obtained for $U > U'$ within the present parameter region. 
However, it seems to express a limitation of the two-site DMFT 
and the consideration in the limit of $U\to\infty$ 
actually leads the following results.

%************** fig.(3) *******************
%\begin{center}
%\end{center}
%******************************************
%************** fig.(4) *******************
%\begin{center}
\begin{figure}[tb]
 \includegraphics[height=5.14cm,clip,keepaspectratio]{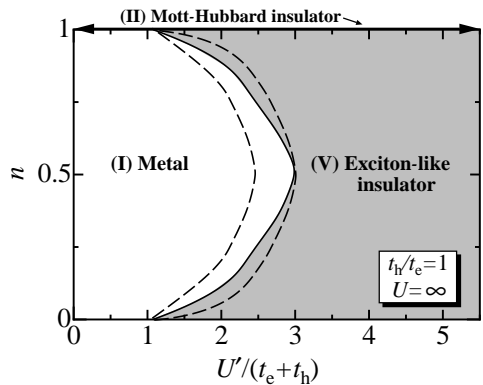}
\caption{%
Phase diagram in the $U'$-$n$ plane 
with $t_\h/t_\e=1$ for the limiting case of large $U$, 
referred from Ref.~\cite{Capone02}. 
The interval area between two dashed curves is the coexistent 
region of the metallic and exciton-like insulating states. 
}%
\label{Fig4}
\end{figure}
%\end{center}
%******************************************

%------------- figure 4 ----------------------
In the limit of $U\to\infty$, the model~(\ref{H})  
 can be mapped onto a single-band attractive Hubbard model
with the attraction $-U'$. 
From the results of the DMFT study 
of  this model~\cite{Keller02,Capone02},  then, 
we can draw the phase diagram on the plane of $U'$ and $n$  
in the limit of $U\to\infty$,  as shown in Fig.~\ref{Fig4}. 
In addition to the metallic state (I) and the Mott-Hubbard insulator 
(II) (just on $n=1$ for all values of $U'$),  
the exciton-like insulating state (V) appears, in which 
the incoherent local e-h pairs (do not condense) 
are formed.  
It is worthy to note that the transition between 
the metallic (I) and exciton-like insulating (V) states 
is the first-order transition (except for $n=0$, 0.5 and 1).   
As well as has been argued in Refs.~\cite{Keller02,Capone02}, 
the exciton-like insulator (V) will be characterized by 
that $Z_\alpha \neq 0$  but $\rho^\alpha(0)=0$ (i.e., 
a gap opens at Fermi level), however 
which cannot be described within the two-site DMFT 
since the gap formation ($V_\alpha=0$) is directly connected to 
$Z_\alpha=0$ through Eq.~(\ref{SCEQ-V}) in the present scheme.  

Let us now return to Fig~\ref{Fig3}. 
Based on both the result of Fig.~\ref{Fig4} 
and the behavior of $Z_\alpha$ in the metallic state, 
it could be surmised that the exciton-like insulator  
appears around the shaded region of Fig.~\ref{Fig3},  
in which $U'$ is comparatively large 
and  $Z_\alpha$ of the metallic solution is quite small.     
To describe the exciton-like insulator 
and determine that phase boundary
for the finite $U$ are left for the future work.  

\vspace{-2mm}
%%%%%%%%%%%%%%%%%%%%%%%%%%%%%%%%%%%%%%%%%%%%%%%%%%%%%
\section{Discussions}
\vspace{-1mm}

In this paper, 
 we found the phase transitions among the exciton-like insulator, 
biexciton-like insulator, and the metallic state, 
for arbitrary fillng with the use of the DMFT. 
This implies that the exciton Mott transition can be 
described essentially in terms of the simple lattice model with 
only short-range interactions
 as well as the Mott-Hubbard transition. 

Finally, we discuss the relevance of our assumption that 
 the e-h pairs do not condense. 
Although the calculation was performed at zero temperature, 
we believe that our present results will be valid 
for the intermediate temperatures, i.e.,   
above critical temperature ($\TC$) of exciton BEC,  but below 
 temperature corresponding to 
the e-h binding energy ($\EB$).   
From simple evaluation of   $\TC$ and $\EB$, 
it can be shown that such a temperature region actually exists: 
consider again in the limit of $U \to \infty$.  
 In the strong limit of $U'$,
$\TC$ can be estimated as 
of order $(t_\e+t_\h)^2/U'$~\cite{Nozieres85,Micnas90}. 
On the other hand, 
in the low-density limit $n \to 0$, 
  $\EB$ can be estimated as of order $U'$~\cite{Micnas90}. 
Comparing these two characteristic temperatures, 
it can be concluded that 
such an intermediate temperature region 
exists even for not so large $U'$ ($\sim t_\e+t_\h$). 
Of course, 
the temperature effect and the problem of condensation 
will be investigated by more precise calculation.

%%%%%%%%%%%%%%%%%%%%%%%%%%%%%%%%%%%%%%%%%%%%%%%%%%%%%
%\section*{Acknowledgements}

This work is supported by CREST, JST.


\vspace{-5mm}
\begin{thebibliography}{00}
%%%=================================================
%--------------
\bibitem{Moskalenko00}
See, for example, 
S. A. Moskalenko, D. W. Snoke, 
Bose-Einstein Condensation of Excitons and Biexcitons,  
Cambridge Univ. Press 2000. 
%--------------
\bibitem{Georges96}
For a review, see A. Georges, G. Kotliar, W. Krauth, M. J. Rozenberg, 
\jo{\RMP}{68}{1996}{13}.
%--------------
\bibitem{Potthoff01}
M. Potthoff, 
\jo{\PRB}{64}{2001}{165114}.
%--------------
\bibitem{Koga02}
A. Koga, Y. Imai, N. Kawakami, 
\jo{\PRB}{66}{2002}{165107}. 
%--------------
\bibitem{Koga04}
A. Koga, N. Kawakami, T. M. Rice, M. Sigrist, 
\jo{\PRL}{92}{2004}{216402}.
%--------------
\bibitem{Keller02}
M. Keller, W. Metzner, U. Schollw\"ock, 
\jo{\JLTP}{126}{2002}{961}.
%--------------
\bibitem{Capone02}
M. Capone, C. Castellani, M. Grilli, 
\jo{\PRL}{88}{2002}{126403}.
%--------------
\bibitem{Nozieres85}
P. Nozi\`eres, S. Schmitt-Rink, 
\jo{\JLTP}{59}{1985}{195}. 
%--------------
\bibitem{Micnas90}
R. Micnas, J. Ranninger, S. Robaszkiewicz, 
\jo{\RMP}{62}{1990}{113}. 
%%%=================================================
\end{thebibliography}
\end{document}